\documentstyle[12pt]{article}
\newcommand \be {\begin{equation}} \newcommand \beq {\begin{equation}}
\newcommand \bea {\begin{eqnarray} \nonumber } \newcommand \ee
{\end{equation}} \newcommand \eeq{\end{equation}} \newcommand \eea
{\end{eqnarray}} \newcommand{\beqa}{\begin{eqnarray}}
\newcommand{\eeqa}{\end{eqnarray}}

\newcommand{\bc}{\begin{center}} \newcommand{\ec}{\end{center}}

 \def\(({\left(} \def\)){\right)} \def\[[{\left[}
 \def\]]{\right]}

\begin{document}                

\begin{titlepage}
\vskip .27in
\begin{center}
{\large \bf Non-equilibrium work relations. }

\vskip .27in

{Jorge Kurchan}
\vskip .2in {\it PMMH-ESPCI, CNRS UMR 7636, 10 rue Vauquelin,
 75005 Paris, FRANCE}
\vskip .2in

\end{center}
\date\today

\vskip 8pt


\vspace{0.5cm}
\begin{center}
{\bf Abstract}
\end{center}
\vspace{0.5cm}

This is a brief  review of  recently derived 
relations describing the 
behaviour of systems far from equilibrium.
They include the
 Fluctuation Theorem,  Jarzynski's and Crooks' 
equalities, and an extended form
of the Second Principle  for general steady states.
They are very general and their proofs are, in most cases,
 disconcertingly simple.

\vspace{1cm}


\end{titlepage}

\vspace{.5cm}
\section{Introduction}
\setcounter{equation}{0}
\renewcommand{\theequation}{\thesection.\arabic{equation}}
\label{Introduction}
\vspace{.5cm}

 In the last few years, a group of relations were derived
involving the distribution of
work done on a system by non-conservative 
and/or time-dependent
forces~\cite{EvCoMo,GaCo,Ja} (for reviews, see \cite{reviews}).
 It seems they have not been noticed before the nineties~\cite{precedents},
even though  in most cases the proofs require techniques that have been
around for many decades. 
I shall discuss them here  in the context of systems that are
    purely Hamiltonian, or,
 when a thermostat is needed, in contact with a stochastic thermal 
bath~\cite{Ku,LeSp}.
In order to stress their similarity, 
before going to the actual proofs, I shall list and comment them: 

\begin{itemize}

\item
 {\em `Transient' Fluctuation Relation~\cite{EvCoMo}}: 
starting from a configuration
 chosen randomly with  the Boltzmann-Gibbs distribution at
 temperature $T=\beta^{-1}$, and applying non-conservative 
 forces  the work done $W$ is a 
random~\footnote{
Even if the dynamics are deterministic, since
the initial condition is itself random.}
variable distributed 
according to a law $P(W)$, which clearly depends upon the process.
We have then the  relation, independent of both model and process:
\begin{equation}
\ln P(W)- \ln 
P(-W)=\beta W
\label{FT}
\end{equation}
An equivalent, and sometimes more direct form is obtained by
multiplying the exponential of 
(\ref{FT}) by $e^{-\lambda W}$ and integrating. For .
real $\lambda$:
\begin{equation}
\langle e^{\lambda W} \rangle_{equil}  = 
\langle e^{-(\beta + \lambda) W} \rangle_{equil}
\label{FT1}
\end{equation}
where $\langle \bullet \rangle_{equil}$ denotes average over trajectories
starting from thermal equilibrium.

\item  {\em Stationary state Fluctuation 
Relation}~\cite{GaCo,comparison,Ku,LeSp}:
In this version, we consider a driven stationary
system. 
 We compute the work
 over a period of time $t$ starting 
from different initial configurations chosen with the stationary,
 (in general non-Gibbsean) distribution.
 A formula like (\ref{FT}) also holds,
but this time 
only for large $t$, when we can write that $P(W)$ converges to
a large-deviation form:
\begin{equation}
\ln P(W) \sim t \zeta\left(\frac{W}{t}\right) + 
\; smaller \;\; orders \;\; in  \;\; t
\end{equation}

\item {\em Jarzynski equality}~\cite{Ja}:
Starting from an equilibrium configuration,
 we vary some parameter $\alpha$
(e.g. magnetic field, volume,...) of the system from
$\alpha_1$ to $\alpha_N$ at an arbitrary rate, and
record the work done $W$.
Averaging over several individual realisations, we have that:
\begin{equation}
\ \langle e^{-\beta W} \rangle_{\alpha_1} = e^{-\beta[
F(\alpha_N)-F(\alpha_1)]}
\label{J}
\end{equation}
Note the puzzling appearance of the {\em equilibrium} free energy at the
final value $\alpha_N$ of the parameter, even though the system is not
assumed to equilibrate at that value!

Clearly, the Jarzynski relation  implies the second principle
\begin{equation}
\langle {\beta W} \rangle \geq \beta[
F(\alpha_N)-F(\alpha_1)]
\label{JJ}
\end{equation}
 through Jensen's inequality $\ln \langle A \rangle 
\geq \langle \ln A \rangle$.

\item {\em Extended Second Law for transitions between non-equilibrium 
stationary  states}~\cite{HaSa}
Consider a driven system with dynamic variables
${\bf x}$ and external fields $\alpha$
(e.g. shear rate, temperature gradient, etc)
that admits non-equilibrium steady states with distribution 
$\rho_{SS}({\bf x};\alpha)=e^{-\phi({\bf {x}};\alpha)}$.
We now take a stationary state at $\alpha=\alpha_1$ and at arbitrary speed
make a transition from  $\alpha_1$ to $\alpha_N$ in such a way that 
stationarity is achieved at $\alpha_N$, but {\em not} necessarily 
at all the intermediate
values.
Repeating the experiment with many initial conditions, the following 
equality holds:
\begin{equation}
\left\langle e^{-\int dt \; \frac{
\partial \phi({\bf {x}};\alpha)
}
{\partial \alpha}
 {\dot{\alpha}}} \right\rangle = 1
\label{HS}
\end{equation}
Again, using Jensen's inequality we obtain:
\begin{equation}
\left\langle \int \frac{
\partial \phi({\bf {x}};\alpha)
}
{\partial \alpha}
d{\alpha} \right\rangle \geq 0
\label{HSP}
\end{equation}
The content of this inequality is that
for a process $\alpha_1,\alpha_2,...,\alpha_N$
 there is a  bound on a functional of the trajectories 
whose form is given entirely by the  values $\alpha$  traversed.
The analogy with the second law is completed by showing that the equality 
holds for a process that can be considered to be in the stationary distribution
at all times, since then:
  \begin{equation}
\left\langle
 \int \frac{
\partial \phi({\bf {x}};\alpha)
}
{\partial \alpha}
 d\alpha 
\right\rangle
 = \int d{\bf{x}} \; d\alpha \; 
 e^{
-\phi({\bf{x}};\alpha)
}
 \frac{
\partial \phi({\bf {x}};\alpha)}{\partial \alpha}=
\int d {\bf{x}}
[\rho({\bf{x}};\alpha_N)-\rho({\bf{x}};\alpha_1)] = 0
\label{HSP1}
\end{equation}
These processes play the role of reversible processes for 
non-driven systems.

In the particular case in which the stationary states for all $\alpha$
are equilibrium states, we have that 
\begin{equation}
\rho_{SS}({\bf x};\alpha) = 
\frac{e^{-\beta E({\bf x},\alpha)}}{e^{-\beta F(\alpha})} \qquad
\phi_{SS}({\bf{x}};\alpha) = \beta (E({\bf x},\alpha) - F^\alpha)
\end{equation}
and (\ref{HS}) and (\ref{HSP}) reduce to a form of the
 Jarzynski equality and the second
law, respectively.

\end{itemize}

\vspace{.5cm}

{\bf NOTES:}

\vspace{.5cm}

$\bullet$ The fluctuation relations are statements about the assymetry
of the  the distribution of work {\em around zero,
and not around the maximum}. They involve  negative work
tails which are usually very rare:
 heat flowing from cold to hot reservoirs,
fluids forcing rheometers to move faster, etc.
These events become  all the more rare  in  macroscopic systems,
since their probability is exponentially small in the size. 
This has 
spurred from the beginning  the search (with some 
success, see \cite{local}) 
 of  `local' versions 
 of the fluctuation relation involving the work on a subsystem, whose 
fluctuations are  more easily observable. 

$\bullet$ Similarly, the average in Jarzynski's equality Eq. (\ref{J})
is dominated by 
 trajectories having a
 very low value of
$W$ which occur with extremely low probability -- exponentially small
in the system size.
This is in strict  analogy with `annealed' averages in 
disordered systems, which are dominated by very
 rare samples with unusually low free energy.

$\bullet$ That the transient 
fluctuation relation and the Jarzynski equality 
are closed relatives can already be seen by considering a process where
initial and final external fields coincide $\alpha_1=\alpha_N$.
In that case $F(\alpha_N)-F(\alpha_1)=0$, and 
 the Jarzynski relation is a consequence of the fluctuation 
relation, since:
\begin{equation}
\langle e^{-\beta W} \rangle =\int dW \; P(W) \; e^{-\beta W}=\int dW \;P(-W)=1
\end{equation}

One can also deduce Jarzynski's relation from the Transient Fluctuation
Theorem by considering the distribution of work done by a {\em conservative}
force switched on at $t=0$, and using the fact that the initial distribution
is Gibbsean.

 A more general connection has been given by Crooks~\cite{Cr},
 who derived a formula
 that contains both. For a system with a time-dependent energy, starting
in equilibrium:
\begin{equation}
\langle {\cal{F}} \rangle_{forward}=\langle {\cal{F}}_r \;
 e^{-\beta W_d} \rangle_{reversed}
\end{equation}
where ${\cal{F}}$ and ${\cal{F}}_r$ are any functional of a trajectory
 and its time-reversed, and $W_d$ is the `dissipative' work --the extra work
with respect to a reversible,  process. 
We obtain the transient fluctuation theorem and the Jarzynski inequalities
putting ${\cal{F}}=\delta\{W-W_d[{\bf x}]\}$ and ${\cal{F}}=1$,
 respectively.

$\bullet$ The fluctuation theorem can also be extended to
cases where the system is driven by the contact with two thermal
baths at different temperatures, rather than by mechanical work~\cite{fourier}.

$\bullet$ The inequality (\ref{HSP1}) is remarkable in that it says that 
there is a bound on a quantity for a process that
is taken from one stationary non-equilibrium state to another,
thus generalising the second law.
The trouble is that, while  for the second law we have an explicit
expression (and a measurement protocol) 
for the work and free energy, we do not have that for
general $\phi(x,\alpha)$.

$\bullet$ In general, these relations are of two different kinds: those
in which the Gibbs-Boltzmann distribution of the initial configuration 
is assumed (transient fluctuation and Jarzynski's relations),
 and those where it is not 
(stationary fluctuation relation, generalised second law).
It turns out that the former are always simple to prove, while  the latter,
though simple for stochastic dynamics, turn out to require more 
sophisticated
tools in deterministic cases. The reason is clear: since we do not assume
equilibration at the outset, some assumption of ergodicity becomes necessary,
and this 
is trivial in systems in contact with stochastic, but not in
 with deterministic baths. 
 In fact, the stationary fluctuation  `Gallavotti-Cohen' theorem\cite{GaCo} 
(valid for a class of deterministic  thermostats) is quite a technical feat,
 and
the Hatano-Sasa generalisation of the second law  for driven systems has not, to
my knowledge, been extended yet to the deterministic case.

$\bullet$ By construction, for the
Langevin systems we shall use here,
we can consider that the heat entering the bath is immediately 
shared by all of it.
This will not be the case with a true (e.g. water) bath,
so that during the process the bath itself falls out of equilibrium,
and in particular has an ill-defined temperature.
 The Jarzynski relation is however still valid,
but one has to include in $\delta F$ also the contribution coming
from the bath-system interaction terms.
See~\cite{CoJa} for a discussion.

$\bullet$ The fluctuation relation is also valid for periodically driven 
systems with symmetric cycles,
both in the transient and in the stationary forms, as can be easily
seen from Crooks' relation.  Here measurements 
are taken at the end of an integer number of cycles, and `stationary' in fact
means
`periodic'.
 
$\bullet$ 
The extension of these results to Quantum systems is rather 
straightforward~\cite{quantum}. In order to avoid the problem of 
introducing non-conservative forces in Shroedinger's equation,
it is sometimes conceptually easier to consider periodic, time-dependent 
 gradient forces.

\pagebreak 

\section{Probability distributions of symmetry-breaking terms.}

As a warming up exercise, let us see what is the effect on the probability 
distribution of a broken discrete symmetry~\cite{spin,Ma}.
Consider a classical system with variables $s_1,...,s_N$ and energy
$E_o({\bf {s}})-\frac{h}{2}M({\bf {s}})$ with $E_o$ having the discrete
 symmetry
$
E_o({\bf {s}})=E_o({\bf {-s}})
$
and 
$
M({\bf {s}})=\sum_i s_i
$
so that 
\begin{equation}
E(-{\bf {s}})= E({\bf {s}})+hM({\bf {s}})
\label{vi}
\end{equation}
and
\begin{equation}
M(-{\bf {s}})= -M({\bf {s}})
\label{sa}
\end{equation}

If $h=0$, the total energy is symmetric, and this 
implies, amongst other things,
 the vanishing of all odd correlation functions.

Can we conclude something in the presence of $h \neq 0$, when the symmetry is 
broken? Indeed, we can: consider the distribution of the symmetry-breaking term
\begin{equation}
P\left[ M({\bf {s}})=-M \right]= \int \; d{\bf{s}}\; 
\delta\left[ M({\bf {s}})+M \right]\;e^{-\beta (E_o-\frac{h}{2}M)}
\end{equation}
Changing variables ${\bf {s}} \rightarrow -{\bf{s}}$, and using the 
symmetry of $E_o$:
\begin{equation}
P\left[ M({\bf{s}})=-M \right]= \int\; d{\bf{s}}\; 
\delta \left[- M({\bf{s}})+M \right]\;e^{-\beta (E_o+\frac{h}M)}=e^{-\beta hM}
P\left [M({\bf{s}})=M \right]
\end{equation}
or
\begin{equation}
\ln P\left[ M({\bf{s}})=M \right]-
\ln P\left [M({\bf{s}})=-M \right]=\beta h M
\label{MM}
\end{equation}
Note that we have said nothing about $E$ apart from the fact that
its symmetry-violating term is $hM$.
An alternative way to state (\ref{MM}) is to say that for any $\lambda$,
\begin{equation}
\langle e^{\lambda M} \rangle =\langle e^{-(\lambda+\beta h) M} \rangle
\label{MM1}
\end{equation}
 where here $\langle \bullet \rangle$ stands for 
equilibrium average.

This relation is clearly valid for the Ising ferromagnet in a field 
in any dimension.
It tells us something about the relation between
the probability of observing a given 
magnetisation and the reversed one, but note that in the thermodynamic limit
one of the two probabilities is vanishingly small.

The formal similarity with the fluctuation relation is obvious.

\pagebreak

\section{Langevin and Kramers Equations}
 
We will consider the Langevin dynamics~\cite{Risken}
\begin{equation}
m {\ddot q}_i + \gamma {\dot q}_i + \partial_{q_i} U({\boldmath q})
+f_i =\Gamma_i,
\label{eq:ksde}
\end{equation} 
where $i=1,\dots,N$.  $\Gamma_i$ is a delta-correlated white noise
with variance $2\gamma T$.
The $f_i$ are velocity-independent forces that do not necessarily
derive from a potential.
The limit $\gamma=0$ corresponds to Hamiltonian
 dynamics.
We shall not consider here the over-damped $m=0$ case, as it brings
about the Ito vs. Stratonovitch convention complications, which add nothing
 conceptually.

If $m \neq 0$ the probability distribution at time $t$ for the process
(\ref{eq:ksde}) is expressed in terms of the phase-space variables
${\bf{x}} \equiv ({\bf{q},{v}})$ and is given by 
\beq \rho({\boldmath q},{\boldmath v},t)=
e^{ -t H} \rho({\boldmath q},{\boldmath v},0) \eeq where $H$ is the
Kramers operator \cite{Risken}:
\begin{equation}
H= \partial_{q_i}v_i - \frac{1}{m} \partial_{v_i} \left (\gamma v_i
+ (\partial_{q_i} U({\boldmath q}))+f_i + \gamma \frac{T}{m}
\partial_{v_i} \right) = H^{c} - \frac{1}{m} \partial_{v_i} f_i
\label{Kramers},
\end{equation}
and $\partial_{q_i}$ and $\partial_{v_i}$ denote derivatives,
and on the r.h.s. we have explicitated the conservative and driven parts.
Denoting $| - \rangle$ with $\langle {\bf{x}}|-\rangle =1$
 the flat distribution, probability conservation is guaranteed by
\begin{equation}
 \langle -| H=0
\end{equation}
A stationary state satisfies:
\begin{equation}
H |\rho_{SS}\rangle =0 \;\;\;\; ; \;\;\;\; \langle -| \rho_{SS}\rangle =1
\end{equation}
In the absence of driving forces:
\begin{equation}
 \langle x| \rho_{SS}\rangle = e^{-\beta (E(x)-F)}
\end{equation}
where $F$ is the free energy at that temperature.

The evolution operator for a dynamics that does not
depend explicitely on time is:
$
U = e^{-tH}
$. For a Langevin process depending upon time through a parameter
(say) $\alpha(t)$, it can be written as a product by dividing 
 $t$ into a large number $M$ of time-intervals: 
\begin{equation}
U = e^{-\frac{t}{M} H(\alpha_M)} ... e^{-\frac{t}{M} H(\alpha_2)}
 e^{-\frac{t}{M} H(\alpha_1)}
\end{equation}
Given an observable $A({\bf{x}};\alpha)$, we can compute an average over
 trajectories, starting from the equilibrium state corresponding
to $\alpha_1$:
\begin{equation}
\langle x| \alpha_1\rangle=\exp\{-\beta(E({\bf{x}};\alpha_1)-F(\alpha_1))\}
\end{equation}
\begin{eqnarray}
\left\langle e^{\int_0^t  dt' 
A({\bf{x}},\alpha(t'))} \right\rangle_{\alpha_1} 
&= \langle -| e^{-\frac{t}{M} H(\alpha_M)}e^{\frac{t}{M} A(\alpha_M) }
 ...
 e^{-\frac{t}{M} H(\alpha_2)}e^{\frac{t}{M}A(\alpha_2) }
 e^{-\frac{t}{M} H(\alpha_1)}e^{\frac{t}{M}A(\alpha_1) } |\alpha_1\rangle
\nonumber \\
\sim & \langle -| e^{-\frac{t}{M} (H(\alpha_M)+A(\alpha_M)) }
 \dots
 e^{-\frac{t}{M} (H(\alpha_2)+A(\alpha_2)) }
 e^{-\frac{t}{M} (H(\alpha_1)+A(\alpha_1)) } |\alpha_1\rangle \nonumber\\
~
\label{martes}
\end{eqnarray}
In the particular case in which $\alpha$ is time-independent:
\begin{equation}
\left\langle e^{\int_0^t  dt' 
A({\bf{x}},\alpha(t'))} \right\rangle_{\alpha} 
= \langle -| e^{-t( H(\alpha)+ A) }|\alpha\rangle
\label{cosas}
\end{equation}

\subsection{ Second Law and reversibility in the stochastic case}

A non-increasing functional of the distribution $\rho({\boldmath x})$
 can be defined as \cite{Kubo}
\begin{equation}
 {\cal{H}}(t)= \int d{\boldmath x}  \rho({\boldmath
x}) \left( T \ln \rho({\boldmath x}) +
E({\boldmath x}) \right)
\label{H}
\end{equation}
and may be interpreted as an out of equilibrium 
 generalised free-energy.
If the energy depends on time via a parameter $\alpha$, a short calculation
using (\ref{Kramers}) gives:
\begin{equation}
\dot{{ {\cal{H}} }} = \int d{\bf{x}} \left[
 \left(\frac{\partial E}{\partial \alpha}
 {\dot{\alpha}}-  {\boldmath f} \cdot {\mathbf v} \right)\rho+
 \gamma \sum_i \int d{\mathbf q} d{\mathbf v} \frac{(mv_i \rho
 -T \partial_{v_i}\rho )^2 }{m^2 \rho } \right]
\label{entr}
\end{equation}
We have, for a closed cycle 
\begin{eqnarray}
& & \oint
 dt \; \left[ \int d{\bf{x}}  \left(
\int d{\bf{x}} \frac{\partial E}{\partial \alpha}
 {\dot{\alpha}}-  {\boldmath f} \cdot {\mathbf v}\right)\rho(x)\right]\nonumber \\
& &=
 \gamma \oint dt 
\sum_i \int d{\mathbf q} d{\mathbf v} \frac{(mv_i \rho
 -T \partial_{v_i}\rho )^2 }{m^2 \rho } \geq 0
\label{second}
\end{eqnarray}
i.e. the `second law' for these systems.

 A reversible process is defined here as 
 {\em  one in which  $\rho({\bf{x}})$ can be considered 
to be at all times Maxwellian in the velocities, so that the r.h.s. vanishes}
\footnote{ In the pure Hamiltonian $\gamma=0$ case 
the equality is satisfied  at all times,
 a consequence of Liouville's theorem.}.
To the extent that Maxwellian distribution of velocities implies the 
Gibbs-Boltzmann distribution in all phase-space variables at all times,
we have that the variation in $ {\cal{H}}$ indeed coincides with the
 equilibrium free-energy difference.
\pagebreak

\subsection{ Time-reversal and work}

The evolution of the system satisfies in the absence of
non-conservative forces a form of detailed balance:
\begin{equation}
 \langle q', v'| e^{-t H^c}| q,v \rangle e^{-\beta E ({\boldmath
 q},{\boldmath v})} = \langle q, -v| e^{-t H^c}| q',- v' \rangle
 e^{-\beta E ({\boldmath q'}, -{\boldmath v}')}
\end{equation}
where the total energy is $E({\boldmath x})=
\frac{1}{2} \sum_i m v_i^2 + U({\boldmath q})$.  This leads to a
symmetry property, which in operator notation reads:
\begin{equation}
Q^{-1} H^c Q = H^{c \; {\dag}}
\label{dbk}
\end{equation}
where the operator $Q$ is defined by:
\begin{equation}
Q |q,v \rangle \equiv e^{-\beta E ({\boldmath x})}
|q,-v \rangle
\label{coco}
\end{equation}

In the presence of arbitrary forces $f_i$, equation (\ref{dbk}) is
modified to:
\begin{equation}
Q^{-1} H Q = H^{\dag} + \beta f_i v_i \equiv H^{\dag} -
S^{\dag} \qquad ; \qquad Q^{-1} (H-S) Q = H^{\dag} 
\label{viernes}
\end{equation}
The operator $S$ is the power done on the system divided by the
temperature \beq S^{\dag} = - \beta {\boldmath f} \cdot {\boldmath
v} \eeq
 We also have:
\begin{equation}
Q^{-1} S Q = - S^{\dag}
\label{sabado}
\end{equation}
 Eqs. (\ref{viernes}) and (\ref{sabado})
are the analogues of equations (\ref{vi}) and (\ref{sa}) above, with 
time-reversal playing the role of spin-flip, and $S$ the role of the 
magnetisation. 
It should come as no surprise that a relation for the large deviations of $S$
is at hand! 

Before closing this section, let us see what the implications of 
(\ref{viernes}) and (\ref{sabado})  
are for a time-dependent evolution.
We start with a product as in (\ref{martes})
\begin{eqnarray}
&  \left[ e^{-\frac{t}{M} H(\alpha_M)}e^{\frac{t}{M}A(\alpha_M) } \dots
 e^{-\frac{t}{M} H(\alpha_2)}e^{\frac{t}{M}A(\alpha_2) }
 e^{-\frac{t}{M} H(\alpha_1)}e^{\frac{t}{M}A(\alpha_1) } \right]^\dag
\nonumber \\
& \; = e^{-\frac{t}{M} (H^\dag(\alpha_1)+A(\alpha_1)) } \dots
 e^{-\frac{t}{M} (H^\dag(\alpha_{M-1})+A(\alpha_2)) }
 e^{-\frac{t}{N} (H^\dag(\alpha_M)+A(\alpha_M)) } 
\label{lunes}
\end{eqnarray}
For each factor, we now use (\ref{viernes}) to transform
the exponents according to:
\begin{equation}
H^\dag(\alpha_l)+A =Q^{-1}(\alpha_l) [ H(\alpha_l) -
 S(\alpha_l)+A_r(\alpha_l)]  Q(\alpha_l) 
\end{equation}
where we had to specify the dependence of $Q(\alpha)$ on $\alpha$ coming
from the dependence of $Q$ on the energy (Eq. (\ref{coco})), and 
we have defined  $A_r({\bf {q},{v}};\alpha) \equiv A({\bf {q},{-v}},\alpha)$.
We hence get:
\begin{eqnarray}
&  \left[ e^{-\frac{t}{M} H(\alpha_M)}e^{\frac{t}{M}A(\alpha_M) }\dots
 e^{-\frac{t}{M} H(\alpha_2)}e^{\frac{t}{M}A(\alpha_2) }
 e^{-\frac{t}{M} H(\alpha_1)}e^{\frac{t}{M}A(\alpha_1) } \right]^\dag 
\nonumber \\
=&  Q^{-1}(\alpha_1)e^{-\frac{t}{M} (H(\alpha_1)-S(\alpha_1)+A_r(\alpha_1)) }
Q(\alpha_1) \dots \nonumber \\
& Q^{-1}(\alpha_{M-1}) e^{-\frac{t}{M}
 (H(\alpha_{M-1})-S(\alpha_{M-1})+A_r(\alpha_{M-1})) } Q(\alpha_{M-1}) \nonumber \\
&  Q^{-1}(\alpha_{M})
 e^{-\frac{t}{N} (H(\alpha_M)-S(\alpha_M)+A_r(\alpha_M)) }  Q(\alpha_{M})
\label{jueves}
\end{eqnarray}
Now,
\begin{equation}
Q^{-1}(\alpha_l)Q(\alpha_{l+1})=e^{-\beta(E(\alpha_{l+1})-E(\alpha_l))}
\end{equation}
and we finally obtain, for large $M$:
\begin{eqnarray}
& & \left[ e^{-\frac{t}{M} H(\alpha_M)}e^{\frac{t}{M}A(\alpha_M) }\dots
 e^{-\frac{t}{M} H(\alpha_2)}e^{\frac{t}{M}A(\alpha_2) }
 e^{-\frac{t}{M} H(\alpha_1)}e^{\frac{t}{M}A(\alpha_1) } \right]^\dag 
\nonumber \\
&=& Q^{-1}(\alpha_1) e^{-\frac{t}{M} (H(\alpha_1)+B(\alpha_1)) } \dots
 e^{-\frac{t}{M} (H(\alpha_{M-1})+B(\alpha_{M-1})) }
 e^{-\frac{t}{N} (H(\alpha_M)+B(\alpha_M)) } Q(\alpha_M) \nonumber \\
& &~
\label{domingo}
\end{eqnarray}
where we have defined:
\begin{equation}
B({\bf {q},{v}};\alpha_{l})= A_r(\alpha_{l})-
\beta[S(\alpha_{l})
-(E(\alpha_{l})-E(\alpha_{l-1})] 
\label{newf}
\end{equation}
This is the general formula for time-reversal, and is in fact two steps
away from both the Fluctuation Theorems and Jarzynski's equality.

\section{Relations}

\vspace{.5cm}

{\em Crooks' Relation}

\vspace{.5cm}

Let us compute:
\begin{eqnarray}
\left\langle e^{\int_0^t  dt' 
A({\bf{x}},\alpha(t'))} \right\rangle_{\alpha_1} 
&=& \langle -| e^{-\frac{t}{M} H(\alpha_M)}e^{\frac{t}{M} A(\alpha_M) }
 ... 
 e^{-\frac{t}{M} H(\alpha_1)}e^{\frac{t}{M}A(\alpha_1) } |\alpha_1\rangle
\nonumber \\
&=& \langle\alpha_1 | \left[e^{-\frac{t}{M} H(\alpha_M)}e^{\frac{t}{M} A(\alpha_M) }
 ... 
 e^{-\frac{t}{M} H(\alpha_1)}e^{\frac{t}{M}A(\alpha_1) }\right]^\dag |-\rangle
\nonumber \\
&=& \langle\alpha_1 |
 Q^{-1}(\alpha_1) e^{-\frac{t}{M} (H(\alpha_1)+B(\alpha_1)) } \dots
 e^{-\frac{t}{N} (H(\alpha_M)+B(\alpha_M)) } Q(\alpha_M) |-\rangle
\nonumber \\
& &~
\label{martes1}
\end{eqnarray}
where we have used (\ref{domingo}). Now, from Eq. (\ref{coco}):
\begin{equation}
 Q(\alpha_M) |-\rangle = |\alpha_M\rangle e^{\beta F(\alpha_M)} \qquad
\; \qquad  \langle\alpha_1 |
 Q^{-1}(\alpha_1) = e^{-\beta F(\alpha_1)}\langle-|
\end{equation}
so that we get:
\begin{eqnarray}
\left\langle e^{\int_0^t  dt' 
A({\bf{x}},\alpha(t'))} \right\rangle_{\alpha_1} 
&=& \langle -| e^{-\frac{t}{M} H(\alpha_M)}e^{\frac{t}{M} A(\alpha_M) }
 ... 
 e^{-\frac{t}{M} H(\alpha_1)}e^{\frac{t}{M}A(\alpha_1) } |\alpha_1\rangle
\nonumber \\
&=&  e^{-\beta (F(\alpha_M)-F(\alpha_1))} \langle- |
  e^{-\frac{t}{M} (H(\alpha_1)+B(\alpha_1)) } \dots
 e^{-\frac{t}{N} (H(\alpha_M)+B(\alpha_M)) } |\alpha_M \rangle
\nonumber \\
&=& \left\langle e^{
\int_0^t  dt' A_r({\bf{x}},\alpha(t')) -\beta W_d } 
\right\rangle_{\alpha_M}
\label{martes2}
\end{eqnarray}
where 
\begin{equation}
W_d= F(\alpha_M)-F(\alpha_1) -\int dt \; \left(
 {\boldmath f} \cdot {\mathbf v}-\frac{\partial E}{\partial t}\right)
\end{equation}
This is Crooks' equality, which relates  averages over two groups
of trajectories, 
starting from equilibrium at $\alpha_1$ and at $\alpha_M$, respectively.

\vspace{.5cm}

{\em Jarzynski's Relation}

\vspace{.5cm}

Putting $A=0$ we immediately get Jarzynski's relation, for a process 
starting in equilibrium for $\alpha_M$:
\begin{equation}
1=\left\langle e^{-
 \beta W_d } 
\right\rangle_{\alpha_M}
\end{equation}

\vspace{.5cm}

{\em Transient Fluctuation Theorem.}

\vspace{.5cm}

To obtain the transient fluctuation theorem for a  system 
with time-independent parameters   ${\dot{\alpha}}=0$ we set:
\begin{equation}
A=\lambda \left(
 {\boldmath f} \cdot {\mathbf v}-\frac{\partial E}{\partial t}\right)
\end{equation}
so that $\int dt\; A=-\int dt\; A_r=W_d$, and
Eq. (\ref{martes2}) becomes, in this case:
\begin{equation}
\left\langle e^{
\lambda W_d} \right\rangle_{\alpha}=
  \left\langle e^{
 -(\lambda +\beta)W_d } 
\right\rangle_{\alpha}
\label{martes3}
\end{equation}
which is the transient Fluctuation Theorem in the form (\ref{FT1}).
Multiplying both sides by $e^{-\lambda W}$ ($W$ a constant)
and integrating both sides over  $\lambda$
we obtain a representation of the $\delta$-function, and we thus have:
\begin{equation}
\left\langle \delta(W_d-W) \right\rangle_{\alpha}=
  \left\langle \delta(W_d+W)  
\right\rangle_{\alpha} e^{\beta W}\qquad ; \qquad
P(W_d=W)=P(W_d=-W)e^{\beta W}
\label{martes5}
\end{equation}
i.e., Eq. (\ref{FT}).

\vspace{.5cm}

{\em Generalised Second Law: the Hatano-Sasa derivation.}

\vspace{.5cm}

Hatano and Sasa have given a truly simple derivation of a 
generalisation of the Second Law, valid for transitions between
non-equilibrium steady states.
Let us denote the steady states associated with a fixed (time-independent) 
value of
the parameter $\alpha$ as $|\rho(\alpha)\rangle$ (and
$ \rho_{SS}(\alpha_a)=\langle {\bf{x}}|\rho(\alpha_a)\rangle$):
\begin{equation}
H(\alpha_a)|\rho(\alpha_a)\rangle=0
\end{equation}
Let us compute for small time steps:
\begin{equation}
\langle-| e^{-\frac{t}{M} H(\alpha_M)} 
\frac{ \rho_{SS}(\alpha_M)}{ \rho_{SS}(\alpha_{M-1})}
... 
\frac{ \rho_{SS}(\alpha_3)}{ \rho_{SS}(\alpha_2)}
e^{-\frac{t}{M} H(\alpha_2)}
\frac{ \rho_{SS}(\alpha_2)}{ \rho_{SS}(\alpha_1)}
 e^{-\frac{t}{M} H(\alpha_1)}|\rho(\alpha_1)\rangle=1
\end{equation}
which can be proved just by going from left to right step by step.
Defining:
\begin{equation}
\phi_{SS}({\bf{x}};\alpha_a) = -\ln \rho_{SS}({\bf{x}};\alpha_a)
\end{equation}
we have  that:
\begin{equation}
\left\langle e^{-
\int dt \; \phi_{ss}({\bf{x}};\alpha) {\dot{\alpha}} } \right
\rangle_{\rho_{SS}(\alpha_1)}=1
\end{equation}
i.e. Eq. (\ref{HS}).

\vspace{.5cm}

{\em The Stationary Fluctuation Theorem}

\vspace{.5cm}

Consider again the transient fluctuation theorem for a system
with  time-independent parameters   ${\dot{\alpha}}=0$.
In looking for a stationary version of the fluctuation relation, 
we shall need to consider the power over long times. In preparation
for this, let us rescale the quantities in (\ref{martes5}) in a natural way.
Denoting $t$ the total time, define  $w=W/t$ as the work per unit time,
$\pi(w)=tP(wt)$ the corresponding probability distribution, and
\begin{equation} 
\zeta(w)\equiv\frac{\ln \pi(w)}{t}
\end{equation}
the logarithm of the probability.
We have then that (\ref{martes5}) becomes:
\begin{equation}
\zeta_\alpha(w)-\zeta_\alpha(-w)= \beta t w
\end{equation}
where we have added the subindex $\alpha$ to remind that the relation is valid
at all times starting from the equilibrium corresponding to $\alpha$.

If we now prove the (very reasonable) result that the distribution
of work per unit times becomes, for large time-intervals, independent of
the initial situation, then this last relation will apply to all cases.
In other words, if we prove that:
\begin{equation}
\zeta_{any}(w)-\zeta_\alpha(w)= {\mbox{ order smaller than }}t  
\label{times}
\end{equation}
then we shall have 
\begin{equation}
\zeta(w)-\zeta(-w)= \beta t w + {\mbox{ order smaller than }}t  
\label{statt}
\end{equation}
for any initial configuration, in particular one obtained from the stationary
(non-Gibbsean) distribution.
{\em Note that the difficulties we might encounter 
are  the usual ones with 
 ergodicity issues, and we are not to blame  the Fluctuation Relation itself}. 
 Here is where the approach with a stochastic bath becomes simpler than the one with a
deterministic bath: ergodicity in the former is automatic, though of course
 put by hand.

Let us start by writing:
\begin{equation}
P(W)=\left\langle \delta(W_d-wt) \right\rangle_{init}=\int d\lambda 
\left\langle e^{
\lambda (W_d-wt)} \right\rangle_{init}=\int d\lambda 
\langle - | e^{
(H+ \lambda S)} |init \rangle e^{-t\lambda w}
\end{equation}
where we have used (\ref{cosas}),
 starting from the distribution $|init\rangle$.

 Introducing the right and left eigenvectors:
\beq (H+ \lambda S) |\psi_a^R(\lambda) \rangle = \mu_a(\lambda)
|\psi_a^R(\lambda) \rangle \;\;\;\; ; \;\;\;\; \langle
\psi_a^L(\lambda)|(H+ \lambda S) = \mu_a (\lambda) \langle
\psi_a^L(\lambda)|
\label{eigenvectors}
\eeq we have:
\begin{equation}
e^{t \zeta_{init}(w)} =  t \int_{-i \infty}^{+i
 \infty} \; d \lambda \; \sum_a \; \langle -|\psi_a^R(\lambda) \rangle
 \langle \psi_a^L (\lambda)|init \rangle e^{-t(\mu_a(\lambda)-\lambda w)}
\end{equation}
Denoting $\mu_0(\lambda)$ the eigenvalue with lowest real part
and the corresponding left and right eigenvectors $ |\psi_0^R(\lambda)
\rangle $ and $ |\psi_0^L(\lambda) \rangle $ and  assuming
 that the eigenvalue $\mu_0(\lambda)$ is non-degenerate,
 the integral over $\lambda $ will be dominated for large $t$ by
the saddle point value:
\begin{equation}
 \zeta(w) = \mu_0(\lambda_{sp})- \lambda_{sp}  w
\label{cosa2}
\end{equation}
where the saddle point $\lambda_{sp}$ is a function of $w$ determined
by:
\begin{equation}
 \frac{d \mu_0}{d\lambda} (\lambda_{sp})=  w
\label{cosa3}
\end{equation}
{\em We have shown that the  dependence upon the initial distribution is
 sub-dominant for large $t$, which is all we needed}.
In doing these last steps, we have assumed that the initial state has non-zero
overlap with the lowest eigenvalue, and that the limit $t \rightarrow \infty$ 
is taken before any other {\em including the pure Hamiltonian }
 $\gamma \rightarrow 0$ one.
All these assumptions 
are just ways of saying that we are assuming that the system 
reaches the same steady state in finite times, a fact
 of which we are guaranteed
for a finite, non-zero temperature stochastic system with bounded potentials.

\vspace{.5cm}

\section{Deterministic  versus Stochastic Dynamics}

\vspace{.5cm}

The results derived in these sections fall in two classes:
those  that do not need (but may have) a thermostat and those for which
 thermostatting is unavoidable.

 The first class includes Crooks' relation
and its consequences the Transient Fluctuation  and the Jarzynski relations.
 These relations hold under the  hypothesis that the system starts from the
Gibbs-Boltzmann measure. The derivations are equally valid in the purely
 deterministic $\gamma=0$ case.
On the contrary, the stationary 
Fluctuation Theorem and the Hatano-Sasa derivation
involve stationary driven states, and this
 requires  thermostatting
(in particular, we can not set $\gamma=0$ in this case).

We have avoided all potential difficulties by working with a stochastic 
Langevin bath, thanks to which the properties of thermostatted
steady states are extremely 
simple.
Had we chosen to work instead with a deterministic thermostat, we would have to
face all the difficulties associated with a fractal stationary   
measure, and the question of chaoticity of the dynamics would also 
become important.
This is what makes the stationary fluctuation Gallavotti-Cohen~\cite{GaCo}
 theorem
quite more involved that anything we have attempted here.

Is the extra effort of working with deterministic thermostats worth it?
From the   point of view of understanding irreversibility and ergodicity 
in general,
 the answer is clearly yes. A very convincing way to see this is to consider
the  derivation leading to `second principle' for systems with
Langevin baths Eq. (\ref{second}): its two-line simplicity is appealing,
but it  teaches us very little about how
 irreversibility
arises in real Hamiltonian systems.

There is however a context in which a stochastic approach is very convenient.
The Stationary Fluctuation Relation has been checked  in a number of
 experiments \cite{reviews}.
 The thermostatting mechanism in such cases is never
  of the (deterministic, time-reversible)  kind assumed in the 
Gallavotti-Cohen theorem,  and we do not know whether the chaoticity properties
assumed in that theorem hold. Indeed, the experiment is sometimes
presented as a test of whether those assumptions are reasonable (if perhaps 
not literally true) for the system in question.
 It is sometimes argued on physical grounds that since the thermostat is
`far away' from the relevant part of system,
 its exact nature
 is irrelevant. Clearly, if 
it is true  that the details of a remote thermostat 
become irrelevant, it is much more convenient 
to consider  a stochastic thermostat, 
for which the  Stationary Fluctuation Relation is trivially and generally 
valid, without any assumptions on the system itself.

\section{Conclusions}

This review  addresses a literature with several 
hundred articles in a few pages, so it is naturally very incomplete.
The most serious omission is the experimental work, and the possible
practical applications -- the reader may find this in
 the reviews~\cite{reviews}.

The literature is quite divided on the role of these relations. In order
of decreasing optimism, 
whether they really teach us something deep about irreversibility, 
or are  a practical tool for measuring properties of medium scale systems,
or are  interesting simply because `there is not much else' that is known far
from equilibrium.  
It seems fair to say that, whatever the final conclusion,
 these equalities give us a more complete perspective of 
the Second Law.

\end{document}